\documentclass[twocolumn,superscriptaddress,amsmath, amssymb, amsfonts,preprintnumbers,aps,prd,longbibliography,nofootinbib]{revtex4-1}

\usepackage{scalerel}
\usepackage{color}
\usepackage{amsmath,amsfonts,amssymb}
\usepackage[small,bf,hang]{caption}
\usepackage{slashed}
\usepackage{latexsym,epsfig}
\usepackage{dsfont}
\usepackage{arydshln}
\usepackage{extarrows}
\usepackage{hyperref}
\usepackage{array}

\def\moth{\mathsurround=0pt}
\newdimen\zo \zo=0pt

\def\tick{\leaders\hrule height 0.5ex depth 0pt \hskip 0.5pt}
\def\upboxfill{$\moth \setbox\zo\hbox{\tick}%
  \hskip 3pt\hbox to 0pt{$\tick$\hss}\hrulefill \hbox to 7.5pt{$\tick$\hss}$}

\def\dtick{\leaders\hrule height .34pt depth 0.5ex \hskip 0.5pt}
\def\downboxfill{$\moth \setbox\zo\hbox{\dtick}%
  \hskip 2pt\hbox to 0pt{$\dtick$\hss}\hrulefill \hbox to 2pt{$\dtick$\hss}$}

\def\ov{\bar}

\def\bec{\begin{center}}
\def\ec{\end{center}}

\def\un{\underline}
\def\ov{\overline}

\def\nn{\nonumber}

\def\be{\begin{equation}}
\def\ee{\end{equation}}
\def\bea{\begin{eqnarray}}
\def\eea{\end{eqnarray}}
\def\ba{\begin{array}}
\def\ea{\end{array}}

\begin{document}

\title{Double Field Theory with matter and the generalized Bergshoeff-de Roo identification}

\author{Eric Lescano} 
\email{elescano@irb.hr}
\affiliation{Division of Theoretical Physics, Rudjer Boskovic Institute, Bijenicka 54, 10000 Zagreb, Croatia}

\author{Nahuel Mirón-Granese}
\email{nahuelmg@fcaglp.unlp.edu.ar}
\affiliation{Consejo Nacional de Investigaciones Científicas y Técnicas (CONICET),
Godoy Cruz 2290, Ciudad de Buenos Aires C1425FQB, Argentina}

\affiliation{ Facultad de Ciencias Astronómicas y Geofísicas, Universidad Nacional de La Plata, Paseo del Bosque, La Plata B1900FWA, Buenos Aires, Argentina }

\affiliation{Departamento de Física, Facultad de Ciencias Exactas y Naturales, Universidad de Buenos Aires,\\Intendente Güiraldes 2160, Ciudad Universitaria, Ciudad de Buenos Aires C1428EGA, Argentina}

\begin{abstract}
The scalar field-perfect fluid (sf-pf) correspondence shows that the energy-momentum tensor of a scalar field is in correspondence with the dynamics of a perfect fluid. In this work we generalize this concept to study the higher-derivative structure of Double Field Theory with statistical matter. Using the generalized Bergshoeff-de Roo identification we find nontrivial higher-derivative corrections for the generalized scalar field Lagrangian. However, these contributions are removed to any desired order using field redefinitions at the supergravity level. By virtue of the generalized sf-pf correspondence we obtain the higher-derivative dynamics for the perfect fluid in the double geometry, which is also trivialized at the supergravity level. These results imply that the well-known $\alpha'$-corrections obtained by this procedure only correct the effective vacuum Lagrangian, while the sf-pf Lagrangian coupled to the supergravity background remains uncorrected to all orders. Our findings apply for a general (Bosonic-Heterotic-Type II-HSZ) supergravity background. 

\end{abstract}

\maketitle

\section{Introduction}

In recent works \cite{HohmZ} a classification for higher-derivative duality invariant theories for purely time-dependent backgrounds was achieved. These theories admit non-perturbative dS solutions in the string frame, and they can be written in a duality invariant way. From a cosmological point of view, these $\alpha'$ contributions correct the vacuum part of the Einstein equations. Matter terms were added later, mimicking the standard formulation of the leading-order string cosmology and extending it to all orders in $\alpha'$ \cite{Brandenmatter}. Within this formalism, if a matter Lagrangian contains higher-derivative terms, a systematic procedure was presented to take rid of all these contributions. When the matter terms come from a statistical energy-momentum tensor, like the energy-momentum tensor of a perfect fluid, the previous method does not apply straightforwardly because of the lack of a matter Lagrangian.

A way to overcome this problem is to use a correspondence between the dynamics of a scalar field and the dynamics of the perfect fluid. Since the former obeys a variational principle, the scalar field-perfect fluid (sf-pf) correspondence can be used to define a fluid Lagrangian \cite{scalarfluidgr}. In here we propose to use the formalism of Double Field Theory (DFT) \cite{Siegel} \cite{DFT} and the generalized Bergshoeff-de Roo identification \cite{gbdr} \cite{gbdr2} to show that the $\alpha'$-corrections for the perfect fluid dynamics are trivial at the supergravity level for arbitrary backgrounds. Consequently the cosmological reductions described in \cite{HohmZ} and \cite{Brandenmatter} contain $\alpha'$ corrections only in their vacuum (or gravitational\footnote{We include dilaton or b-field contributions in here.}) part of their dynamics.

Since T-duality is an exact symmetry of string theory \cite{Ashoke}, its supergravity limit can be cast in a duality invariant way. The standard procedure consists of a regrouping of the fundamental fields in $O(D,D)$ multiplets \footnote{We use $D$ for the critical dimension of the string theory formulation, and we focus on the universal NS-NS (Neveu-Schwarz) sector.}, while the geometry has to be modified in a suitable way and, for example, the dimensions of the space must be doubled (standard DFT formulation). One way to extend DFT for including statistical matter contributions is to construct a double kinetic theory \cite{phaseDFT}. Within this framework, the conservation law for the generalized current of particles and the energy-momentum tensor can be derived in agreement with the variational principle of DFT \cite{ParkCosmo}. In \cite{cosmoDFT} the explicit T-duality covariant form of the generalized energy-momentum tensor was given for a perfect fluid in the double geometry. Furthermore, the dynamics of this extension of DFT is given by string cosmologies with dilaton charge when one imposes cosmological ansatz for the generalized metric/dilaton, and upon imposing strong constraint and field redefinitions for the DFT energy density and pressure. The construction given in \cite{cosmoDFT} suggests a formal correspondence between a generalized scalar field and a perfect fluid in the double geometry.

The main goal of this work is to show that higher-derivative terms induced by the generalized Bergshoeff-de Roo identification applied to a generic DFT background with a perfect fluid coupled only correct the vacuum dynamics upon parametrization. Since the starting point of this mechanism requires an $O(D+K,D+K)$ DFT, having a zeroth-order matter Lagrangian for the perfect fluid is mandatory. For this reason we first explore the sf-pf correspondence at the supergravity level, which can be use to construct a variational principle for the latter. Then we generalize this correspondence at the DFT level obtaining a variational principle for the perfect fluid in the double geometry. We use the generalized Bergshoeff-de Roo identification coupling this Lagrangian to an $O(D+K,D+K)$ invariant DFT and we obtain the higher-derivative deformation of DFT in presence of a perfect fluid. Finally, we parametrize the theory and take a field redefinition to obtain an invariant metric tensor under Lorentz transformations, which in turn trivialize all the higher-derivative terms in the matter Lagrangian. We highlight three novel results in this work:
\begin{enumerate}
\item  We obtain the explicit correspondence between a generalized scalar field and the DFT perfect fluid. The procedure consists in inspecting the generalized energy-momentum tensor in each case and to obtain a formal relation between the derivatives of the scalar field and the hydrodynamics variables of the fluid in the double geometry. The generalized correspondence results in a variational principle for the DFT perfect fluid, where the Lagrangian is given by the DFT pressure, ${\cal L}_{m}=\tilde p$. Moreover, this result implies that the energy-momentum tensor given in \cite{cosmoDFT} obeys a generalized version of the correspondence. The matter Lagrangian for the perfect fluid in the double geometry is proportional to the DFT pressure and it reproduces the standard matter Lagrangian obtained by the ordinary correspondence after parametrization.

\item We couple the sf-pf Lagrangian to an $O(D+K,D+K)$ extended geometry and perform the generalized Bergshoeff-de Roo identification. After a gauge fixing procedure, which consists in identifying the gauge fields with the $O(D+K,D+K)$ fluxes considering the matter contributions, we obtain higher-derivative corrections in terms of $O(D,D)$ fields. The new contributions for the matter sector deform the generalized sf-pf Lagrangian in the double geometry, and they are induced because of the anomalous transformation of the generalized metric under double Lorentz transformations.

\item At the supergravity level, only the vacuum Lagrangian receives higher-derivative corrections when a sf-pf is coupled: the DFT corrections for the matter are removed to any desired order using a metric field redefinition. Our results explain from a geometrical point of view the lack of $\alpha'$ corrections for the matter in cosmological backgrounds, as reported in a previous work \cite{Brandenmatter}. However, our procedure is valid for generic $a$ and $b$ values of the generalized Bergshoeff-de Roo identification and therefore it applies to generic supergravity backgrounds, in particular Bosonic-Heterotic-Type II-HSZ supergravity among others.

\end{enumerate}

\section{The scalar-fluid correspondence in a supergravity background}

The scalar-fluid correspondence establishes a formal identification between a minimally coupled scalar field and an effective perfect fluid \cite{scalarfluidgr}. This correspondence is very useful in order to include an effective matter Lagrangian for the latter.

Let us focus on the supergravity action with matter
\bea
S= \frac12 \int d^{D}x\,e^{-2 \varphi}\, \sqrt{-g} \, \left[R+4 (\partial \varphi)^2 - \frac{1}{12} H^2 \right] + S_{\textrm{mat}} \nn
\eea
where
\bea
S_{\textrm{mat}} = \int d^{D}x\,e^{-2 \varphi}\, \sqrt{-g} \,L_{\rm mat}\,.\label{smat}
\eea
The equation of motion for the metric tensor is given by
\bea
&& G_{\mu \nu} + 2 \nabla_\mu \nabla_\nu \varphi+ 2g_{\mu \nu}(\nabla\varphi)^2 -2g_{\mu \nu}\nabla^2\varphi  \nn \\ && + \frac{1}{24} g_{\mu \nu} H^2 - \frac{1}{4} H_{\mu \rho \sigma} H_{\nu}{}^{\rho \sigma} = e^{2\varphi}\,T_{\mu \nu} \, , 
\label{emt}
\eea
here $\mu,\nu=0,\dots,D-1$, $G_{\mu \nu}$ is the well-known Einstein tensor, $G_{\mu\nu}=R_{\mu \nu} - \frac12 R g_{\mu \nu}$, $T_{\mu \nu}$ is the standard energy-momentum tensor coming from the matter contribution
\bea
T_{\mu \nu}=\frac{-2}{\sqrt{-g}} \frac{\delta S_{\rm mat}}{\delta g^{\mu \nu}}\,,\label{deftmunu}
\eea
and the exponential factor $e^{2\varphi}$ acts as an effective gravitational coupling. The dynamics of the dilaton field reads
\bea
R - 4 (\partial \varphi)^2 - \frac{1}{12} H^2  + 4 \nabla_{\mu}(\nabla^{\mu} \varphi) =- e^{2\varphi}\,\sigma\label{dilatoneq}
\eea
with the dilaton source
\bea
\sigma = \frac{-1}{\sqrt{-g}}\frac{\delta S_{\rm mat}}{\delta\varphi} =- e^{-2\varphi} \left[\frac{\delta L_{\textrm{mat}}}{\delta \varphi} - 2 L_{\textrm{mat}}\right] \, \label{defsigma}
\eea
while the dynamics for the b
 field is given by
\bea
-\frac12\nabla^\rho H_{\rho\mu\nu}+(\nabla^\rho \varphi)H_{\rho\mu\nu}=2e^{2\varphi}J_{\mu\nu}\label{jeq}
\eea
with the b-source
\bea
J_{\mu\nu} = - \frac{2}{\sqrt{-g}}\frac{\delta S_{\rm mat}}{\delta b_{\mu\nu}}\,.\label{defj}
\eea

In the case of a scalar field the Lagrangian for the matter action (\ref{smat}) is
\bea
L_{\rm mat}=-\frac12 \partial^{\mu} \Phi \partial_{\mu} \Phi - V(\Phi)\label{lmatsugra}
\eea
and the dynamics is given by the Klein-Gordon equation 
 \bea
 \Box \Phi - \frac{\delta V}{\delta \Phi}= 0\,.
 \eea
The energy-momentum tensor from (\ref{deftmunu}) is
\bea
T^{\mu \nu}_{\Phi} = e^{-2\varphi}\left[\partial^{\mu} \Phi \partial^{\nu} \Phi - \frac12 g^{\mu \nu} \partial^{\rho} \Phi \partial_{\rho} \Phi - V g^{\mu \nu} \right]\, ,
\label{scalar}
\eea
while the dilaton source (\ref{defsigma}) and the b-source (\ref{defj}) for the scalar field read
\bea
\sigma_{\Phi} &=& 2 e^{-2\varphi}\,\left[-\frac12 \partial^{\mu} \Phi \partial_{\mu} \Phi - V(\Phi)\right] \, , \label{sigmascalarfield}\\
J^{\mu\nu}_{\Phi}&=&0\,,\label{jscalarfield}
\eea
respectively.

On the other hand, the energy-momentum tensor for an effective perfect fluid reads
\bea
T^{\mu \nu}_{\rm PF} = (e+p) u^{\mu} u^{\nu} + p g^{\mu \nu} \, ,
\label{fluid}
\eea
with $e$ the energy density, $p$ the pressure and $u_\mu$ the velocity of the fluid. As usual in cosmological approaches, the starting point for this kind of matter is given by the energy-momentum tensor due to its statistical nature, instead of a matter Lagrangian. If in addition to (\ref{fluid}) we also consider arbitrary sources for the dilaton $\sigma_{\rm PF}$ and the b field $J_{\rm PF}^{\mu\nu}$, we can derive string cosmology equations from (\ref{emt}), (\ref{dilatoneq}), (\ref{jeq}) and (\ref{fluid}) after imposing a cosmological ansatz. Indeed, the independent equations read \cite{Gasperini,Chouha}
\bea
\label{parametrization1}
- e^{2\varphi}\sigma_{\rm PF} &= &  2 (D-1) \dot{H} + D(D-1) H^{2} \nn \\ && -4 \ddot{\varphi} + 4 \dot{\varphi}^2 - 4 (D-1) H \dot{\varphi} \\ 
-e^{2\varphi}\left(e+\frac{\sigma_{\rm PF}}{2}\right)& =&  (D-1) \left(\dot{H} + H^{2}\right) -2\ddot{\varphi} \label{parametrization2}\\
e^{2\varphi}\left( p-\frac{\sigma_{\rm PF}}{2}\right)  & = &  \dot H + (D-1) H^{2}  - 2 H \dot{\varphi}\label{parametrization3}\\
-2e^{2\varphi}J^{\rm PF}_{ij}  & = & \ddot b_{ij}+\left[(D-5)H-2\dot\varphi\right] \dot b_{ij}
\label{parametrization4}
\eea

At the GR level, the scalar-fluid correspondence was studied in \cite{scalarfluidgr} through the formal comparison between their energy-momentum tensors. In this supergravity scenario, an analogous procedure can be carried out in order to achieve the correspondence by comparing the expressions (\ref{scalar}) and (\ref{fluid}). We first define the identification of the velocity as
\bea
u_{\mu} = \frac{\partial_{\mu} \Phi}{\sqrt{|\partial^{\rho} \Phi \partial_{\rho} \Phi|}} \, ,
\eea
with $\partial^{\rho} \Phi \partial_{\rho} \Phi \neq 0$ and $u_{\mu} u^{\mu}=\textrm{sign} (\partial^{\rho} \Phi \partial_{\rho} \Phi)$. Since we use the positive signature, when $\partial^{\rho} \Phi \partial_{\rho} \Phi<0$ the velocity $u_\mu$ defines a time-like vector and the energy density and the pressure of the effective perfect fluid are related to the scalar field through
\bea
e & = & e^{-2\varphi}\left[- \frac12 \partial^{\rho} \Phi \partial_{\rho} \Phi + V \right]\, , \\
p & = & e^{-2\varphi}\left[- \frac12 \partial^{\rho} \Phi \partial_{\rho} \Phi - V \right]\, .\label{psugra}
\eea

Moreover, through this correspondence we can also obtain the sources of the dilaton $\sigma_{\rm PF}$ and the b field $J^{\mu\nu}_{\rm PF}$ in terms of the scalar field as
\bea
\sigma_{\rm PF}&=&2 e^{-2\varphi}\,\left[-\frac12 \partial^{\mu} \Phi \partial_{\mu} \Phi - V(\Phi)\right]\\
J^{\mu\nu}_{\rm PF}&=&0\,,
\eea
which, in fact, implies that $\sigma_{{\rm PF}}= 2p$.

Finally, using (\ref{lmatsugra}) and (\ref{psugra}) we are able to find the matter Lagrangian corresponding to a perfect fluid in terms of its own dynamical variables as 
\bea
L_{\rm mat,\,PF}=e^{2\varphi} \,p\,.\label{lmatpfsugra}
\eea
In the next section we show that this correspondence can also be achieved in the context of DFT.

\section{The generalized scalar-fluid correspondence}
\label{Correspondence}

\subsection{The perfect fluid in the double geometry}

String cosmologies can be rewritten in terms of $O(D,D)$ multiplets before compactification \cite{cosmoDFT}. While the cosmological principle would suggest an invariant rewriting in terms of an ordinary time and a dual one, matter contributions can be defined doubling the full set of coordinates. Since the fundamental dimension of $O(D,D)$ is $2D$, the geometry of DFT consists of a double space with coordinates $X=(\tilde x, x)$, equipped with an invariant group metric and its inverse. Doubled space-time vectors can be defined considering generalized infinitesimal diffeomorphisms, and these transformations are regulated through a generalized Lie derivative,
\bea
{\cal L}_\xi V_M(X) = && \xi^{N} \partial_N V_M(X) + (\partial_M \xi^N - \partial^N \xi_{M}) V_N(X) \nn \\ && + \omega (\partial_{N} \xi^{N}) V_{M}(X) \, ,
\label{glieintro}
\eea 
where $V_M(X)$ is a generic vector with weight $\omega$ on the double space and $M,N,\dots$ are indices in the fundamental of $O(D,D)$. The strong constraint,
\bea
\partial_{M} (\partial^{M} \star) & = & 0 \,  \quad (\partial_{M} \star) (\partial^{M} \star) =  0 \, ,
\label{SC}
\eea
ensures the closure of the generalized diffeomorphisms in terms of a C-bracket and, effectively, removes the extra coordinates $\tilde x$ when the constraints are solved through $\tilde \partial = 0$. The DFT indices are contracted with the invariant group metric 
\bea
{\eta}_{{M N}}  = \left(\begin{matrix}0&\delta_\mu^\nu\\ 
\delta^\mu_\nu&0 \end{matrix}\right)\, ,  \label{etaintro}
\eea
where $\mu,\nu=0,\dots,D-1$. The standard formulation of DFT takes into account vacuum fields known as the generalized metric, ${\cal H}_{M N}$ which is an element of $O(D,D)$, and the generalized dilaton, $d$. For later use we define in here the DFT projectors: $P_{M N} = \frac12(\eta - {\cal H})_{M N}$ and $ \ov P_{M N} = \frac12(\eta + {\cal H})_{M N}$ such that an arbitrary vector can be projected in this way:
\bea
V^{M} = P^{M}{}_{N} V^{N} + \ov P^{M}{}_{N} V^{N} = V^{\un M} + V^{\ov M} \, .
\label{projcurve}
\eea

One can introduce statistical matter in the double geometry \cite{phaseDFT}. In order to describe this kind of matter, the standard construction of DFT needs to be promoted to a double kinetic theory approach. The double phase space of DFT consists on an extension of the double geometry including both the usual doubled coordinates and the doubled momentum coordinates as $(X,\cal P)$, where ${\cal P}=(\tilde p,p)$. Considering a generalized one-particle distribution function $F(X,{\cal P})$, which is a double phase space scalar, the generalized energy-momentum tensor for the statistical matter can be constructed in the following way   
\bea
{\cal T}^{ M  N}( X) = \int d^{2D}{\cal P}\,e^{-2 d} \,
 {\cal P}^{ M} {\cal P}^{ N}  \, F(X,{\cal P})  \, ,\label{emtintro}
\eea
where $e^{-2d}$ is the DFT measure, $\mathcal T_{MN}$ is clearly symmetric and it can be shown that it is also divergenceless. 

When we couple a generalized perfect fluid to the double geometry, the generalized energy-momentum tensor takes the form \cite{cosmoDFT} , 
\bea
{\cal T}_{ M  N} = -2(\tilde e+\tilde p)  \left[U_{\un { M}}  U_{\ov { N}} +  U_{\ov M}  U_{\un { N}}\right] + \tilde p \, {\cal H}_{ M  N} \, ,
\label{emintro}
\eea
where $\tilde e( X)=\tilde e$ and $\tilde p( X)=\tilde p$ are the DFT energy density and pressure and $ U^{ M}$ is a generalized velocity whose parametrization reads $({u}_{\mu}, \tilde u^{\mu})$. Since the parametrization of the generalized velocity is different from the one in \cite{cosmoDFT}, we must also add a negative sign in the first term of (\ref{emintro}) in order to match string cosmology \cite{Gasperini}. 

The expression (\ref{emintro}) reminds the structure of the ordinary energy-momentum tensor in GR, but in DFT we need to impose $U_{M} \eta^{M N} U_{N}=0$ to take rid of the dual velocity $\tilde u_{\mu}$. Another difference between the DFT tensor with respect to its GR version is the presence of mixed projections in the terms which depend on the generalized velocity. In the next subsection we will show that this is a feature of the generalized scalar field Lagrangian, which motivates the existence of a pf-sf correspondence in the double geometry. As expected the energy-momentum tensor (\ref{emintro}) encodes the matter contributions of the string cosmology equations with fixed dilaton charge $\sigma_{\rm PF}=2p$ and a vanishing b-field source $J_{\rm PF}^{\mu\nu}=0$. The procedure for obtaining these equations is to consider a cosmological ansatz for the DFT fields, to solve the strong constraint and then parametrize the fields. The field redefinition $\tilde e=e^{2\varphi}e$ and $\tilde p=e^{2 \varphi}p$ is also needed to match (\ref{parametrization1})-(\ref{parametrization4}).

\subsection{The correspondence in the double geometry}

We start by considering the following action principle,
\bea
{\cal S} & = & \int d^{2D}X e^{-2d}\, {\cal L}\left[{\cal H},d,\Phi \right] \nn \\ & = & \int d^{2D}X e^{-2d} \Bigg(\frac1{2}{\cal R}\left[{\cal H},d\right] +  {\cal L}_{m}\left[{\cal H},d,\Phi\right] \Bigg) \, ,
\label{actionDFT}
\eea
where ${\cal L}_{m}$ represents matter coupled to the background field content and $\Phi$ represents matter fields. The standard way of computing the generalized energy-momentum is using a variational principle such that
\bea
\delta_\xi {\cal S}&=&\int d^{2D}X \left[\frac{\delta \left(e^{-2d}{\cal L}\right)}{\delta d} \delta_\xi d+\frac{\delta \left(e^{-2d}{\cal L}\right)}{\delta {\cal H}^{MN}} \delta_\xi {\cal H}^{MN} \right. \nn \\ & & \left. +\frac{\delta \left(e^{-2d}{\cal L}\right)}{\delta \Phi} \delta_\xi \Phi\right]\nn\\
&=&\int d^{2D}X e^{-2d} \left[\xi^N\nabla^M G_{MN}-\xi^N\nabla^M T_{MN}+\frac{\delta {\cal L}_m}{\delta \Phi}\delta_\xi \Phi\right] \nn
\label{diffeoinvariance}
\eea
where
\bea
G_{MN} =-\frac12\eta_{MN}{\cal R}+2\left({\cal R}_{\overline M\underline N}-{\cal R}_{\underline M\overline N}\right)\label{GMNlagrangian}
\eea
and
\bea
T_{MN} &= & 2\,\Big[\overline{P}_{MK} P_{NL}-\overline{P}_{NK} P_{ML}\Big]\left(\frac{\delta {\cal L}_m}{\delta {P}_{KL}}-\frac{\delta {\cal L}_m}{\delta \overline{P}_{KL}}\right)\nn \\ && + \eta_{MN}\left({\cal L}_m-\frac12\frac{\delta {\cal L}_m}{\delta d}\right)
\label{TMNlagrangian}\,.
\eea
The generalized tensors $G_{MN}$ and $T_{MN}$ are used to construct the generalized symmetric Einstein tensor and the generalized symmetric energy momentum tensor in the following way
\bea
{\cal G}_{MN}=G_{M P}{{\cal H}^{P}}_N\quad {\rm and}\quad {\cal T}_{MN}=T_{MP}{{\cal H}^{P}}_N.
\eea
Moreover, it is possible to summarize the dynamics defining a generalized Einstein equation ${\cal G}_{MN}={\cal T}_{MN}$ from the equations of motion.

At this point it is important to observe that since the generalized dilaton $d$ is a fundamental field which takes part in the DFT measure, the construction of the generalized energy-momentum tensor ${\cal T}^{MN}$ not only includes the contribution of the generalized metric sources but also the generalized dilaton charge, which explains from a variational principle why the perfect fluid in the double geometry dictates the existence of a nonvanishing fixed dilaton source at the supergravity level.

Now we focus on an $O(D,D)$ invariant scalar field coupled to the background content of DFT. We consider that the  DFT matter Lagrangian is given by,
\bea
{\cal L}_{\textrm{matter}}[ {\cal H}, \Phi] = - \frac12  {\cal H}^{ M  N} \partial_{ M}  \Phi \partial_{ N}  \Phi - V(\Phi)\, .
\label{Scalarlm}
\eea
The equation of motion of the generalized scalar field is the equivalent of the Klein-Gordon equation,
\bea
 {\cal H}^{ M  N} \nabla_{ M} \nabla_{ N}  \Phi - \frac{\delta V}{\delta \Phi}= 0 \,
\label{KGDFT}
\eea
and its generalized energy-momentum tensor for the previous field reads
\bea
 {\cal T}_{ M  N} &= & - 4\, \overline{P}_{ K( M}\,P_{ N) L}\,\partial^{ K}  \Phi \partial^{ L} \Phi - \frac12 {\cal H}_{ M  N}  {\cal H}^{ R  Q} \partial_{ R} \Phi \partial_{ Q} \Phi \nn \\&& - {\cal H}_{ MN}\,V(\Phi)\, .
\label{ScalarTMN}
\eea

We first establish the generalization of the fluid velocity and scalar field derivative correspondence as
\bea
 U_{ M} = \frac{\partial_{ M}  \Phi}{\sqrt{|{\cal H}^{PQ} \partial_{P}\Phi \partial_{Q} \Phi}|} \, .\label{velocitytoscalarfield}
\eea
A comparison between (\ref{emintro}) and (\ref{ScalarTMN}) through (\ref{velocitytoscalarfield}) shows that the generalized pressure and energy density are given by
\bea
\label{pcorrespondence}
\tilde p & = & - \frac12 {\cal H}^{PQ} \partial_{P}\Phi \partial_{Q} \Phi - V(\Phi) \, , \\ 
\tilde e + \tilde p & = & |{\cal H}^{PQ} \partial_{P}\Phi \partial_{Q} \Phi| \, ,
\eea
and hence the matter Lagrangian reads
\bea
{\cal L}_{\rm m}=\tilde p\,.
\eea

Using this correspondence the DFT action, when a perfect fluid is coupled to the double geometry, is given by
\bea
{\cal S}_{\textrm{DFT}} = \int d^{2D}X e^{-2d} \Bigg(\frac1{2}{\cal R}\left[{\cal H},d\right] +  \tilde p(X) \Bigg) \, .
\label{action}
\eea
Then the supergravity matter action (eqs. (\ref{smat}) and (\ref{lmatpfsugra})) is recovered when $\tilde p(X) \rightarrow \tilde p(x)$ upon parametrization, namely
\bea
d^{2D}Xe^{-2d}&\rightarrow& d^Dx\, \sqrt{-g}\,e^{-2\varphi} \nn \\
\frac1{2}{\cal R}\left[{\cal H},d\right] +  \tilde p(X) &\rightarrow& \frac1{2}(R+4(\partial \varphi)^2 - \frac{1}{12} H^2) + \tilde p(x) \, , \nn
\eea
and $\tilde p = e^{2 \varphi} p$ with $p$ the physical pressure.

\section{The generalized Bergshoeff-de Roo identification}

\subsection{The generalized Bergshoeff-de Roo identification for the vacuum case}
The higher-derivative extension of the ordinary DFT formulation requires to abandon the generalized metric formalism and to consider a generalized frame, $E_{M}{}^{A}$, as a fundamental field with $A=0,\dots,2D-1$ a double flat index. The frame formulation of DFT is invariant under a local double Lorentz group given by $O(9,1)_L\times O(1, 9)_R$, infinitesimally generated by a parameter $\Lambda_{AB}$. The frame formulation of DFT demands the existence of two constant, symmetric and invariant metrics $\eta_{AB}$ and ${H}_{AB}$. The former is used to raise and lower the flat indices and the latter is constrained to satisfy 
\bea
{H}_{A}{}^{C}{H}_{C}{}^{B} = \delta_{A}^{B}\, .
\eea 
The generalized frame is constrained to relate the metrics $\eta_{{MN}}$ and $\eta_{ {AB}}$ and defines the generalized background metric ${\cal H}_{MN}$ from ${H}_{AB}$ as
\be
\eta_{MN} = E_{M}{}^{A}\eta_{AB}E_{N}{}^{B}\, , \quad {\cal H}_{MN} = E_{M}{}^{A}{H}_{AB}E_{N}{}^{B} \, .
\ee
At this point it is convenient to introduce the flat projectors $P_{AB}=\frac12(\eta - H)_{AB}$ and $\ov P_{AB}=\frac12(\eta + H)_{AB}$. 

The vacuum DFT action principle with higher-derivative terms is given by \cite{gbdr2},
\bea
S = \int d^{2D}X e^{-2d} \left[\frac1{2} {\cal R}(E,d) + \frac1{2} {\cal R}^{(m,n)}(E,d)\right]  \, ,
\label{hdaction}
\eea
where ${\cal R}$ is the generalized Ricci scalar and ${\cal R}^{(m,n)}$ is a higher-derivative Lagrangian which depends on two parameters, $a$ and $b$, such that each term scales as $a^{n} b^{m}$. The action (\ref{hdaction}) is obtained after gauge fixing the zeroth order $O(D+K,D+K)$ invariant DFT and performing the identification of the gauge degrees of freedom in order to obtain a $O(D,D)$ invariant DFT with only gravitational fields. For instance \footnote{For a pedagogical introduction to this topic see \cite{Lect}.}, the ${}^{MN}$ component of $O(D+K,D+K)$ metric encodes an $O(D,D)$ metric plus higher-derivative contributions, 
\bea
{\cal H}^{M N}_{O(D+K,D+K)}=({\cal H} + {\cal I}^{(m,n)})^{M N}_{O(D,D)} \label{hplusi}
\eea
and, notably, the RHS is (double) Lorentz invariant. The terms ${\cal I}_{M N}$ are given in terms of the generalized fluxes of DFT \cite{Exploring},
\bea
{\cal I}_{M N} & = & - a F_{\ov M}{}^{\un A \un B} F_{\ov N \un A \un B} - b F_{\un M}{}^{\ov A \ov B} F_{\un N \ov A \ov B} \nn\\
&&+ {\cal O}(a^2,ab,b^2) \, . \, \label{iexpression}
\eea 

The $a$ and $b$ parameters are present in the Lorentz transformations of the generalized frame which emulate a generalized Green-Schwarz mechanism
\bea
\delta_{\Lambda} E_{M}{}^{A} = && E_{M}{}^{B} \Lambda_{B}{}^{A} + a \,\partial_{[\un P} \Lambda^{\un B \un C} F_{\ov M] \un B \un C} E^{P A} \nn \\ && - b \,\partial_{[\ov P} \Lambda^{\ov B \ov C} F_{\un M] \ov B \ov C} E^{P A} + {\cal O}(a^2,ab,b^2) \, .
\eea
Consequently the Lorentz transformation for the $O(D,D)$ generalized metric is nonvanishing when higher-derivative terms are considered, namely
\bea
\delta_\Lambda \mathcal H_{MN}&= & -2a\,F_{(\ov N|\un{AB}}\,\partial_{\un M)}\Lambda^{\un{AB}} - 2b\,F_{(\un N| \ov{AB}}\,\partial_{\ov M)}\Lambda^{\ov{AB}} \nn\\
&&+ {\cal O}(a^2,ab,b^2) \,.\label{metrictransffirstorder}
\eea

When the $O(D,D)$ metric ${\cal H}_{MN}$ is parametrized using a symmetric and antisymmetric tensor $\Big\{ \tilde g_{\mu \nu}, \tilde b_{\mu \nu} \Big\}$ a field redefinition for the symmetric tensor is required to obtain a Lorentz invariant (inverse) metric $g^{\mu \nu}$,
\bea
g^{\mu \nu} = && \tilde g^{\mu \nu} + \Delta \tilde g^{\mu \nu} \, .
\label{redefg}
\eea
The form of the field redefinition $\Delta \tilde g^{\mu \nu}$ can be easily obtained taking $\Delta \tilde g^{\mu \nu}={\cal I}^{(m,n)\mu \nu}$. With this choice it is possible to trivialize the inverse of the metric tensor to any desired order and then it is straightforward to trivialize the metric. Hence the anomalous transformation of $\tilde g_{\mu \nu}$ is removed, meaning that the metric tensor $g_{\mu \nu}$ is Lorentz invariant. Once the metric redefinition is imposed, the dilaton is also redefined according to $e^{-2d}=\sqrt{- \tilde g} e^{-2 \tilde \phi} = \sqrt{-g} e^{-2 \phi}$. We do not comment on the redefinition of the b field since it will play no role in the matter Lagrangian.

\subsection{Coupling the scalar field}
The action principle which describes an scalar field $\Phi$ coupled to the $O(D+K,D+K)$ geometry is given by 
\bea
{\cal L}_{m} = \left[-\frac12{\cal H}^{-1} \partial \Phi \partial \Phi - V(\Phi)\right]_{O(D+K,D+K)} \, .
\label{MatterKK}
\eea
This Lagrangian reduces to the following $O(D,D)$ invariant matter Lagrangian (with higher-derivative terms) according to the generalized Bergshoeff-de Roo identification,
\bea
{\cal L}_{m}[ E, \Phi] = && - \frac12  ({\cal H} + {\cal I}^{(m,n)})^{M N}  \partial_{ M}  \Phi \partial_{ N}  \Phi - V(\Phi)\, .
\label{scalarfieldlagrangian}
\eea
Since (\ref{MatterKK}) contains a pair of $O(D+K,D+K)$ derivatives acting on the generalized scalar field contracted with the $O(D+K,D+K)$ metric the ${}^{MN}$ component is the only relevant contribution, i.e. we only need the expression (\ref{hplusi}) to construct the higher-derivative matter Lagrangian (\ref{scalarfieldlagrangian})\footnote{Here we are using the fact that the generalized partial derivatives in the gauge-directions vanish.}. Therefore the higher-derivative action principle for this configuration reads
\bea
{\cal S} = \int && d^{2D}X e^{-2d} \Bigg( \frac1{2}{\cal R}\left[{\cal H},d\right] + \frac1{2} {\cal R}^{(m,n)}(E,d)) \nn  \\ && - \frac12  ({\cal H} + {\cal I}^{(m,n)})^{M N}  \partial_{ M}  \Phi \partial_{ N}  \Phi - V(\Phi) \Bigg) \, .
\eea

The parametrization of the matter contribution in the previous action principle is
\bea
- \frac12  ({\cal H} + {\cal I}^{(m,n)})^{\mu \nu}  \partial_{\mu}  \Phi \partial_{\nu}  \Phi - V(\Phi)\nn\\
=- \frac12  (\tilde g + \Delta \tilde g)^{\mu \nu}  \partial_{\mu}  \Phi \partial_{\nu}  \Phi - V(\Phi)\, .
\eea
where we use $\partial_M=(\tilde \partial^\mu,\partial_\mu)$ with $\tilde\partial=0$ given by the usual solution to the strong constraint. By virtue of the mandatory redefinition (\ref{redefg}), the higher-derivative corrections of the scalar field Lagrangian are removed to any desired order. Hence, when one couples a scalar field to the double geometry, only the background fields receive higher-derivative corrections in their dynamics upon parametrization. In the next part of the section we take advantage of this result for the scalar field and we use it, together with the sf-pf correspondence at the DFT level, to study the higher-derivative structure of perfect fluid dynamics.

\subsection{Coupling the perfect fluid}

The perfect fluid Lagrangian can be directly coupled to the $O(D,D)$ geometry with higher-derivative terms promoting the correspondence (\ref{pcorrespondence}) to $O(D+K,D+K)$. This procedure produces the following action principle,
\bea
{\cal S} = \int d^{2D}X e^{-2d} &\Bigg[ & \frac1{2}{\cal R}\left[{E},d\right] + \frac1{2} {\cal R}^{(m,n)}\left[E,d\right] \nn \\ && +  \big( \tilde p + \tilde p^{(m,n)} \big) \Bigg] \, .
\eea
 where $\tilde p^{(m,n)} =  - \frac12 {\cal I}^{PQ} \partial_{P}\Phi \partial_{Q} \Phi  \, .$ 
 
 Upon parametrization and considering the field redefinition (\ref{redefg}) the perfect fluid matter Lagrangian recovers its leading order form, just as happens in the parametrization of the generalized scalar field Lagrangian,
\bea
\tilde p + \tilde p^{(m,n)} \rightarrow e^{-2 \varphi} p \, . 
\eea

The sf-pf dynamics have a rich and nontrivial higher-derivative interactions at the DFT level which deform, for instance, the algebraic $L_{\infty}$ structure of the theory \cite{Linf}. These interactions are a genuine deformation of the matter Lagrangian and cannot be absorb using field redefinitions of the generalized metric. Nonetheless, it turns out that, after parametrization, these corrections are trivialized for all $a$ and $b$ at any desired order. At this point it is important to recall that the generalized Bergshoeff-de Roo identification recovers all the $\alpha'$ corrections for the Heterotic and Bosonic supergravity up to $\alpha'^2$ \cite{gbdr2,LNR,lastWulff} for particular values of the parameters $a$ and $b$, but new deformations of the model are required to include $\alpha'^3$ contributions as pointed out in \cite{Wulff}.

\section{Discussion}

Most of the string cosmologies scenarios in the literature are based on the coupling of a perfect fluid to a cosmological supergravity background \cite{Brandenmatter, ParkCosmo, cosmoDFT, Gasperini, Chouha}. In some cases, the matter Lagrangian is represented by the Helmholtz free energy, which is equivalent to the pressure Lagrangian here discussed for a canonical ensemble of particles \cite{vafa} after integration of the spatial volume. These kind of cosmological ansatz can be easily obtained from the double geometry as showed in \cite{cosmoDFT} paying the cost of a nonvanishing fixed dilaton charge. The latter is mandatory if one wants to preserve the DFT measure which indeed ensures the invariance under generalized diffeomorphisms.

In this work we show that the perfect fluid dynamics coupled to a generic supergravity background does not receive higher-derivative terms on its dynamics. Considering the generalized version of the sf-pf correspondence our result is related to the families of covariant field redefinitions discussed in \cite{Walter}. In the context of supergravity with $O(d,d)$-invariant matter, the absence of $\alpha'$ corrections was also noticed in \cite{Brandenmatter} for an arbitrary matter Lagrangian within the cosmological framework. The authors used the formalism developed in \cite{HohmZ} in order to get rid of the higher-derivative contributions, in agreement with the results presented here for general backgrounds. In other words, the generalized version of the sf-pf correspondence makes a formal analogy between the generalized scalar physics and the dynamics of a perfect fluid in the double geometry and since the higher-derivative corrections of the former can be trivialized at the supergravity level so the corrections for the fluid at the same level. As a particular case, the cosmological framework is not corrected with $\alpha'$ corrections or other higher-derivative terms which can be put in a T-duality invariant way as in HSZ theory \cite{HSZ}.  

On the other hand, the results of this work show that the proposal for the generalized the energy-momentum tensor given in \cite{cosmoDFT} implies a formal correspondence between the generalized scalar field and the fluid in the double geometry. Our results explains why it is mandatory a dilaton charge in order to find agreement between DFT and string cosmologies. Furthermore, it is expected that the generalized correspondence can be a useful tool to explicitly construct this tensor from (double) kinetic theory \cite{phaseDFT}, where the form of the generalized distribution function is not known.  

Deforming the perfect fluid structure could be a way to include nontrivial higher-derivative terms coming from the matter Lagrangian coupled to a DFT background. The current obstruction of this program relies in the fact that this deformation would include nonvanishing entropy production and the entropy current is not well-understood, at the DFT level, when statistical matter is present in the double geometry. These imperfect fluids may require a new correspondence in order to use the generalized Bergshoeff-de Roo identification.

\section{Conclusions}
\label{Conclusions}

We explore the scalar-fluid correspondence for a generic D-dimensional supergravity background.
On the one hand we couple a scalar field and, on the other, we consider a perfect fluid with generic dilaton and b-field sources. The structure of the energy-momentum tensors for both types of matter allows a formal correspondence between them. Consequently it is possible to find a matter Lagrangian for the perfect fluid which is indeed proportional to the pressure.

We generalize this correspondence for a generic DFT background, where the vacuum fields are given by the generalized metric/dilaton and the matter is coupled considering a generalized scalar field and a double perfect fluid. The generalized version of the sf-pf correspondence allows us to construct a variational principle for the latter, whose Lagrangian is proportional to the DFT pressure. Upon parametrization, this pressure can be related to the ordinary pressure up to a field redefinition.

The higher-derivative structure of DFT with matter is analyzed using the generalized Bergshoeff-de Roo identification. Interestingly enough, the identification produces nontrivial higher-derivative corrections for the generalized sf-pf Lagrangian. However these corrections are removed to any desired order considering a field redefinition for the metric tensor at the supergravity level. These results are valid for generic D-dimensional supergravity backgrounds and explain the absence of $\alpha'$ corrections for the perfect fluid dynamics in the cosmological ansatz.

\subsection*{Acknowledgements}
We thank
D. Marqués, W. Barón, S. Hronek and the IGFAE group (Instituto Galego de Física de altas enerxías) for interesting discussions. E.L is very grateful to Universidade de Santiago de Compostela for hospitality during the last stage of this work. N.M.G is supported by CONICET PIP 11220170100817CO grant, while E.L is supported by the Croatian Science Foundation project IP-2019-04-4168.


\begin{thebibliography}{}
\bibitem{HohmZ}
O. Hohm and B. Zwiebach, 
`Non-perturbative de Sitter vacua via $\alpha'$ corrections, Int. J. Mod. Phys. D 28 (2019) 14, 1943002, [hep-th/1905.06583].

O. Hohm and B. Zwiebach, `Duality invariant cosmology
to all orders in $\alpha'$', Phys. Rev. D 100, 126011 (2019),
[hep-th/1905.06963].

C. A. Nunez and F. E. Rost, `New non-perturbative de Sitter vacua in $\alpha'$-complete cosmology', JHEP 03 (2021) 007, [hep-th/2011.10091]. 

T. Codina, O. Hohm, D. Marques, `General String Cosmologies at Order $\alpha'^3$', Phys. Rev. D 104 (2021) 10, 106007, [hep-th/2107.00053].

T. Codina, O. Hohm, D. Marques, `An $\alpha'$-complete theory of cosmology and its tensionless limit', [hep-th/2211.09757].

\bibitem{Brandenmatter}
H. Bernardo, R. Brandenberger, G. Franzmann, `$O(d,d)$ covariant string cosmology to all orders in $\alpha'$', JHEP 02 (2020) 178, [hep-th/1911.00088].

\bibitem{scalarfluidgr}
V. Faraoni, `Correspondence between a scalar field and an effective perfect fluid', Phys. Rev. D {\bf 85} (2012) 024040, [gr-qc/1201.1448].

\bibitem{Siegel}
  W.~Siegel, `Two vierbein formalism for string inspired axionic gravity', Phys.\ Rev.\ D {\bf 47} (1993) 5453, [hep-th/9302036]. 

W.~Siegel, `Superspace duality in low-energy superstrings', Phys.\ Rev.\ D {\bf 48} (1993) 2826, [hep-th/9305073].

W. ~Siegel, `Manifest duality in low-energy superstrings', In *Berkeley 1993, Proceedings, Strings ’93* 353-363, and State U. New York Stony Brook - ITP-SB-93-050 (93,rec.Sep.) 11 p. (315661), [hep-th/9308133].

\bibitem{DFT}
 C.~Hull and B.~Zwiebach, `Double Field Theory', JHEP {\bf 09} (2009) 099, [hep-th/0904.4664].

  O.~Hohm, C.~Hull and B.~Zwiebach,
  `Generalized metric formulation of Double Field Theory',
  JHEP {\bf 08} (2010) 008,
  [hep-th/1006.4823].

J. H. Park, `Comments on double field theory and diffeomorphisms', JHEP {\bf 06}
(2013) 098, [hep-th/1304.5946].

 D. S. Berman, M. Cederwall, and M. J. Perry, `Global aspects of double geometry',
JHEP {\bf 09} (2014) 066, [hep-th/1401.1311].

I.~Jeon, K.~Lee and J.~H.~Park,
  `Stringy differential geometry, beyond Riemann',
  Phys.\ Rev.\ D {\bf 84} (2011) 044022,
  [hep-th/1105.6294].

  I.~Jeon, K.~Lee and J.~H.~Park, `Differential geometry with a projection: Application to Double Field Theory',
  JHEP {\bf 04} (2011) 014,
  [hep-th/1011.1324].



\bibitem{gbdr}
W.~H.~Baron, E.~Lescano and D.~Marques,
  `The generalized Bergshoeff-de Roo identification',
  JHEP {\bf 1811}, 160 (2018), [hep-th/1810.01427].

\bibitem{gbdr2}
W. Baron and D. Marques, ``The generalized Bergshoeff-de Roo identification. Part II", JHEP 01 (2021) 171, [hep-th/2009.07291] 

\bibitem{Ashoke}
A.~Sen, `$O(d) \times O(d)$ symmetry of the space of cosmological solutions in string
theory, scale factor duality and two-dimensional black holes', Phys. Lett. B {\bf 271} (1991) 295.

\bibitem{phaseDFT}
E. Lescano and N. Mirón-Granese, `On the phase space in Double Field Theory', JHEP 07 (2020) 239, [hep-th/2003.09588].

\bibitem{ParkCosmo}
S. Angus, K. Cho and J. H. Park, `Einstein Double Field Equations', Eur. Phys. J. C {\bf 78} (2018) 500, [hep-th/1804.00964].

S. Angus, K. Cho, G. Franzmann, S. Mukohyama and J.-H. Park, `$O(D,D)$ completion of
the Friedmann equations', Eur. Phys. J. C \textbf{80} (2020) 830, [hep-th/1905.03620].

\bibitem{cosmoDFT}
E. Lescano and N. Mirón-Granese, `Double Field Theory with matter and its cosmological application', [hep-th/2111.03682].

\bibitem{Gasperini}
M. Gasperini, `Elements of
string cosmology', Cambridge University Press, 2007, ISBN:978-0-521-18798-5.

\bibitem{Chouha}
H. Bernardo, P. R. Chouha, and G. Franzmann, `Kalb-Ramond backgrounds in $\alpha$'-complete cosmology', JHEP \textbf{09} (2021) 109 [hep-th/2104.15131]

\bibitem{Exploring}
D.~Geissbuhler, D.~Marques, C.~Nunez and V.~Penas,
  `Exploring Double Field Theory,'
  JHEP {\bf 1306} (2013) 101, [hep-th/1304.1472].


\bibitem{LNR}
E. Lescano, C.A. Nuñez and A. Rodríguez, ``Supersymmetry, T-duality and Heterotic $\alpha'$-corrections", [hep-th/2104.09545].

\bibitem{Lect}
E. Lescano,`$\alpha'$-corrections and their double formulation', [hep-th/2108.12246]. 

\bibitem{Wulff}
S. Hronek and L. Wulff, `$O(D,D)$ and the string $\alpha'$ expansion: an obstruction', JHEP 04 (2021) 013, [hep-th/2012.13410]

\bibitem{lastWulff} 
S. Hronek, L. Wulff and S. Zacarias, `The $\alpha'^2$ correction from double field theory', [hep-th/2206.10640]

\bibitem{Linf}
  O.~Hohm and B.~Zwiebach,
  `$L_{\infty}$ Algebras and Field Theory',
  Fortsch.\ Phys.\  {\bf 65} (2017) no.3-4,  1700014, [hep-th/1701.08824].
  
  E. Lescano and M. Mayo, `Gauged double field theory as an $L_{\infty}$ algebra', JHEP 06 (2021) 058, [hep-th/2103.07361].  

\bibitem{vafa} A. A. Tseytlin and C. Vafa, `Elements of string cosmology', Nucl. Phys. B 372 (1992) [hep-th/9109048].

\bibitem{Walter}
W. H. Baron, `Duality covariant field redefinitions', Phys. Rev. D 105 (2022) 10, 106015, [hep-th/2201.00030] 

\bibitem{HSZ}
 O.~Hohm, W.~Siegel and B.~Zwiebach,
  ``Doubled $\alpha'$-geometry'',
  JHEP {\bf 1402}, 065 (2014),
  [hep-th/1306.2970].

E.~Lescano and D.~Marques,
  ``Second order higher-derivative corrections in Double Field Theory'',
  JHEP {\bf 1706}, 104 (2017), [hep-th/1611.05031].




\end{thebibliography}
\end{document}